\documentclass[]{aa}  

\usepackage{natbib}
\bibpunct{(}{)}{;}{a}{}{,}
\usepackage{graphicx}
\usepackage{revsymb}	
\usepackage{amsmath}	
\usepackage[usenames]{color}	
\usepackage{epstopdf}	
\usepackage{txfonts}
\usepackage{amssymb}
\usepackage{color}
\usepackage[justification=centering]{caption}
\usepackage{geometry} 
\usepackage[parfill]{parskip} 
\usepackage{amssymb}
\usepackage{slantsc}
\usepackage{array}
\usepackage{url}
\usepackage{amsfonts}
\usepackage[colorlinks=true,linkcolor=blue, urlcolor=blue, citecolor=blue]{hyperref}
\usepackage{sidecap}
\usepackage{graphicx}
\usepackage{xcolor}
\usepackage{nicefrac}
\usepackage{subfigure}
\usepackage{epsfig}
\colorlet{rouge}{red!70!darkgray}

\begin{document}
\title{Investigating the impact of Solar Fusion III reaction rates on helioseismic constraints and solar neutrino fluxes}
\author{T. Sandron\inst{1} \and G. Buldgen\inst{1} \and A. Noels\inst{1} \and M.A. Dupret\inst{1} \and R. Scuflaire\inst{1}}
\institute{STAR Institute, Université de Liège, Liège, Belgium \\email: \texttt{t.sandron@outlook.com}}
\date{January, 2026}
\abstract{Nuclear reaction rates are crucial ingredients of solar and stellar models, they directly impact the duration of the life of stars and the energy they produce. In the solar case, the tight observational constraints put on models (Mass, Radius, Luminosity and chemical composition) coupled to our capabilities to probe the solar interior thanks to helioseismology and solar neutrinos provide an exquisite testbed for such physical ingredients. With the recent publication of the Solar Fusion III reaction rates, a new generation of solar models may be computed and put to the test.}
{We aim to investigate the impact of the new Solar Fusion III reaction rates on solar models, both standard and non-standard, as well as the impact of the current uncertainties on some key solar reactions on the predicted neutrino fluxes for boron and the so-called CNO cycle.}
{We compute various theoretical standard solar models as well as non-standard models reproducing the depletion of lithium and beryllium for various abundances, nuclear reaction rates and opacities. We focus on the impact of the solar fusion III reaction rates on both helioseismic inversion results and neutrino fluxes.}
{We find that using the Solar Fusion III reaction rates significantly impact the agreement of solar models both with helioseismic constraints and neutrino flux measurements. While for helioseismic constraints it seems that there is a slight improvement, for neutrino fluxes, the use of the SFIII reaction rates induces a lowering of the beryllium, boron and CNO neutrino fluxes. When investigating the impact of changes on the rates of key reactions within their quoted uncertainties, we find that these changes are far from sufficient to reconcile models and observations, potentially hinting at other processes (e.g. planetary formation as found in other studies).}{}
\titlerunning{The impact of Solar Fusion III reaction rates on solar models} \authorrunning{T. Sandron et al.}
\maketitle
\section{Introduction}
Nuclear reaction rates are key elements of solar and stellar evolution models. They determine the duration of the various evolutionary phases of the Sun and stars. In the solar case, they also influence the agreement of solar models with helioseismic constraints and solar neutrino fluxes. The Sun is a primary testbed for physical ingredients of stellar models, thanks to the precise determination of the position of the base of the convective envelope \citep{KosovBCZ,JCD91Conv}, the helium mass fraction in the convective zone \citep[see e.g.][]{Vorontsov1991,RichardY,DiMauro2002,BasuY2004} as well as our ability to scan through the solar structure using helioseismic constraints and the precise measurements of neutrino fluxes providing direct insights on the conditions of the deep solar core \citep{Bahcall1989,Haxton2013}. This allows to test in great details the impact of varying physical ingredients of solar models, providing key insights on their accuracy \citep[see][for a detailed review]{JCD2021}. In the era of space-based photometry missions, the question of physical ingredients of stellar evolution models is also deeply tied to the accuracy and precision of the inferences of stellar mass, radius and age, a key objective of the upcoming PLATO mission \citep{Rauer2025}. 

Over the years, various compilations of nuclear reaction rates have been produced \citep{Angulo1999,Adelberger2011,Xu2013} with some specific revisions of key rates made independently, such as that of the so-called bottleneck reaction of the CNO cycle, the protonic capture of nitrogen to produce oxygen, noted in short $^{14}\rm{N}(\rm{p},\gamma)^{15}\rm{O}$ \citep{Formicola2004,Imbriani2005}. In parallel, renewed experimental efforts aim at improving the precision of the measured reaction rates, providing crucial data for solar and stellar evolutionary models \citep[e.g.][]{Marta2008,Broggini2010,Marta2011,Wagner2018,Chen2025}. In this study, we investigate the impact of the recently published Solar Fusion III (SFIII) reaction rates \citep{Acharya2025} on solar models, using both helioseismic and neutrino constraints. To achieve this goal, we compute a large set of solar calibrated models, both within the Standard Solar Model framework \citep{Bahcall1968,Bahcall1982,Bahcall1988,JCD1996} and including macroscopic transport. 

We start in Sect. \ref{Sec:Models} by presenting the set of solar models we will use in our study. In Sect. \ref{Sec:HelioModels} and \ref{Sec:NeutrinosModels}, we study the agreement of a first subset of models computed using various abundances, opacities, modified electronic screening and the SFIII reaction rates with helioseismic and neutrino constraints. In Sect. \ref{Sec:UncertaintyNuclear}, we discuss the existing uncertainties on key reactions influencing the neutrino fluxes, namely the $^{7}\rm{Be}(\rm{p}, \gamma)^{8}\rm{B}$ and the $^{14}\rm{N}(\rm{p},\gamma)^{15}\rm{O}$ reactions that influence both the so-called boron neutrino flux, $\phi_{\rm{B}}$ and the CNO neutrino flux, $\phi_{\rm{CNO}}$. In Sect. \ref{Sec:Discussion}, we discuss in more details the observed differences stemming from the revision of the nuclear reaction rates and how they compare to the existing quoted uncertainties. We then discuss potential sources of improvements to restore the agreement between solar models and the observed neutrino fluxes.

\section{Solar Evolutionary Models}\label{Sec:Models}

We start by presenting in Table \ref{tabModels} the solar evolutionary models calibrated in this study. We provide all the key elements of the models relevant for neutrino flux computations. We considered both standard (SSM) and non-standard solar models (NSSM) with additional macroscopic mixing at the base of the convective zone (BCZ) calibrated to reproduce the solar lithium and beryllium depletions \citep{Wang2021,Amarsi2024}.  

Each model is computed with the Liège Stellar Evolution Code \citep[CLES][]{ScuflaireCles} using the Solar Fusion III reaction rates from \citet{Acharya2025}, except for two Standard Solar Models (SSM) who were computed with a combination of the NACRE II and Solar Fusion II reaction rates used in our previous works \citep{Adelberger2011,Xu2013} to provide comparison points. All models were computed using version 7 of the SAHA-S equation of state and the OPAL opacities \citep{OPAL}, except one SSM computed using the updated Los Alamos opacities \citep{Colgan}. All models include microscopic diffusion following \citet{Thoul} and using the screening coefficients of \citet{Paquette}. Non-Standard Solar Models (NSSM) include a parametric turbulent diffusion coefficient at the base of the convective zone (BCZ) following \citet{Proffitt1991}. Convection was treated using the Mixing-Length Theory following the implementation presented in \cite{CoxGiuli1968} and included an atmosphere model following Model-C of \citet{Vernazza}. We also tested models where we deactivated the electronic screening to emulate the effects of dynamical screening that might impact the solar calibration \citep{Shaviv2001,Mao2009,Mussack2011A,Mussack2011B,Dappen2024}. This was done by simply putting to one the screening coefficients of all the reactions, both in the pp chains and in the CNO cycle. The models with modified reaction rates were based on reported uncertainties in the literature and will be discussed in Sect. \ref{Sec:UncertaintyNuclear}.

Solar calibrations were computed using the initial hydrogen mass fraction, $\rm{X_{0}}$, the initial heavy element mass fraction, $\rm{Z_{0}}$ and the mixing-length parameter, $\alpha_{\rm{MLT}}$ as free parameters. The constraints used were the solar radius, $\rm{R}_{\odot}$, the solar bolometric luminosity, $\rm{L}_{\odot}$ and the solar metallicity $\rm{(Z/X)_{S}}$. The two former constraints were fixed to the recommended IAU values \citep{Mamajek2015} and for the latter, we adopted the \citet{Asplund2021} abundances (hereafter AAG21) as our main references but also tested the results for the \citet{Magg2022} abundances (hereafter MB22).

\begin{table*}[h]
\caption{Global parameters of the solar evolutionary models.}
\label{tabModels}
  \centering
\begin{tabular}{r | c | c | c | c | c | c}
\hline \hline
\textbf{Name}&\textbf{$\rm{X}_{\rm{C}}$}&\textbf{$\rm{Z}_{\rm{C}}$}& $\mathit{\rho}_{\rm{C}}$ ($\rm{g/cm^{3}}$) & $\mathit{T}_{\rm{C}}$ ($\rm{(K)}$)& \textbf{$\rm{r}_{\rm{CZ}}$ $(R_{\odot})$} & \textbf{$\rm{Y}_{\rm{CZ}}$}\\ \hline
SSM AAG21&$0.3602$&$0.01642$& $148.4527$& $15.4408\times 10^{6}$ & $0.7219$& $0.2379$\\
SSM MB22&$0.3477$&$0.01945$& $149.5492$ & $15.5762\times 10^{6}$ & $0.7163$& $0.2466$\\
SSM MB22 OPLIB&$0.3551$&$0.01956$& $152.9150$ & $15.4094\times 10^{6}$ & $0.7141$& $0.2411$ \\ 
SSM AAG21 (NACRE II + SFII)&$0.3604$&$0.01647$& $149.6878$& $15.4845\times 10^{6}$ & $0.7229$ & $0.2372$\\
SSM MB22 (NACRE II + SFII)&$0.3476$&$0.01950$& $150.7914$ & $15.6207\times 10^{6}$ & $0.7173$ & $0.2459$\\ 
NSSM AAG21&$0.3653$&$0.01567$& $147.9121$ & $15.3855\times 10^{6}$ & $0.7252$ & $0.2456$\\ 
NSSM MB22&$0.3528$&$0.01863$& $149.1246$ & $15.5201\times 10^{6}$ & $0.7191$& $0.2534$\\
NSSM AAG21 Screening Off&$0.3676$&$0.01571$& $151.0298$& $15.4765\times 10^{6}$ & $0.7281$ & $0.2421$\\
NSSM MB22 Screening Off&$0.3551$&$0.01869$& $152.2101$ & $15.6099\times 10^{6}$ & $0.7218$ & $0.2525$\\
NSSM AAG21 Be$^{7}$(3.41$\%)$&$0.3653$&$0.01567$& $147.9118$& $15.3855\times 10^{6}$ & $0.7252$ & $0.2456$\\
NSSM AAG21 Be$^{7}$(10$\%)$&$0.3653$&$0.01567$& $147.9121$ & $15.3854\times 10^{6}$ & $0.7252$ & $0.2456$ \\
NSSM AAG21 Be$^{7}$(3.41$\%$) N$^{14}$(30$\%$)&$0.3646$&$0.01566$& $147.9615$& $15.3895\times 10^{6}$ & $0.7252$ & $0.2456$ \\
\hline
\end{tabular}
\tablefoot{The central hydrogen and metal mass fraction: $\rm{X}_{\rm{C}}$ and $\rm{Z}_{\rm{C}}$; the central density and temperature: $\mathit{\rho}_{\rm{C}}$ and $\mathit{T}_{\rm{C}}$; the position of the BCZ and the helium mass fraction in the CZ: $\rm{r}_{\rm{CZ}}$ and $\rm{Y}_{\rm{CZ}}$.}
\end{table*}

As can be seen from Table \ref{tabModels}, using the SFIII nuclear reaction rates has a significant impact on the global properties of solar models. The core density, $\rho_{\rm{C}}$, is reduced by $\approx 0.6\%$, while the core temperature is only reduced by $\approx 0.3\%$. The surface helium mass fraction is also significantly affected, as well as the BCZ position. The observed changes in Table \ref{tabModels} may appear small, but at the level of precision of helioseismic and neutrino constraints, we will see in the following section that they are not negligible. 

\subsection{Helioseismic constraints}\label{Sec:HelioModels}

In this section, we present helioseismic inversion results for some of the models presented in Table \ref{tabModels}, namely of the squared adiabatic sound speed profile, $c^{2}=\frac{\Gamma_{1}\rm{P}}{\rho}$, and entropy proxy profile \citep{Buldgen2025S}, $S=\frac{\rm{P}}{\rho^{\Gamma_{1}}}$, with P the pressure, $\rho$ the density and $\Gamma_{1}$ the first adiabatic exponent. The inversions were computed using the SOLA method \citep{Pijpers} by solving the variational equations \citep{Dziembowski91} for both sound speed and entropy proxy. The surface correction and trade-off parameters were adjusted following \citet{RabelloParam}. We used $\approx 2200$ individual frequencies, taken from \citet{BasuSun} and \citet{Davies}.

\begin{figure*}
	\centering
		\includegraphics[width=15cm]{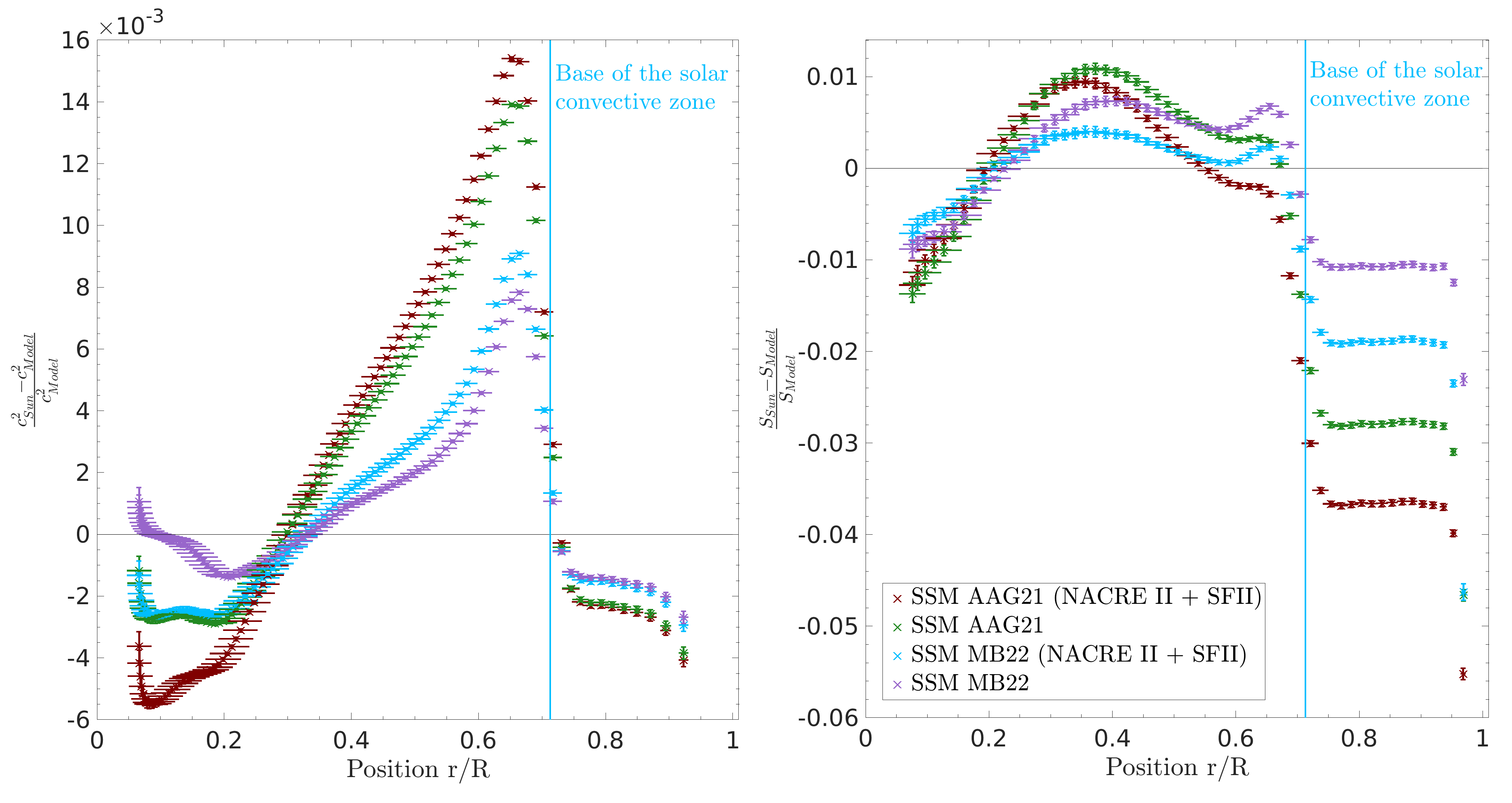}
	\caption{Left panel: Relative differences in squared adiabatic sound speed profile $(\rm{c}^{2})$ between the Sun and various SSMs of Table \ref{tabModels} as a function of normalized radius, as determined from an SOLA inversion of helioseismic data. Right panel: Relative differences in entropy proxy profile $(\rm{S})$ between the Sun and various SSMs as a function of normalized radius, as determined from an SOLA inversion of helioseismic data.}
		\label{Fig:C2SSM}
\end{figure*}

We start in Fig. \ref{Fig:C2SSM} with SSMs using both the AAG21 and MB22 abundances combined with either the SFIII nuclear reaction rates or our previous compilations from NACRE II and SFII presented in previous papers. As mentioned above, the variations induced by the change of nuclear reaction rates are not negligible at all at the level of precision of helioseismic constraints. Using the SFIII reaction rates improves the overall agreement of the sound speed profile, illustrated in the left panel of Fig. \ref{Fig:C2SSM}, in the core regions by less than $0.2\%$, with more modest improvements being also observed at the BCZ and in the envelope. Such modifications are a result of the repercussions of the lower core density that have to be compensated in the rest of the solar structure, given that both the mass and the radius of the model are known and reproduced with high precision in the calibration procedure. 

The situation is also clear from the entropy proxy inversion, as this effect on the density profile directly impacts the height of the entropy proxy plateau. From the right panel of Fig. \ref{Fig:C2SSM}, we can see that the differences reduce significantly in the convective zone (CZ) when using the SFIII reaction rates, as the lower core density impacts the rest of the structure. However, the agreement is worsened on the radiative side of the BCZ when compared with the models using the NACRE II and SFII reactions. This effect is observed whatever the abundances, as the situation is exactly the same with models using the MB22 abundances, with the discrepancies in the core being this time slightly larger than for models using the AAG21 abundances. The effects are reminiscent of the switch from the OPAL to the OPLIB opacities illustrated in \citet{BuldgenS,Buldgen2017A,Buldgen2019} while the effect on the core density, as shown in Table \ref{tabModels}, are the exact opposite.

\begin{figure*}
	\centering
		\includegraphics[width=15cm]{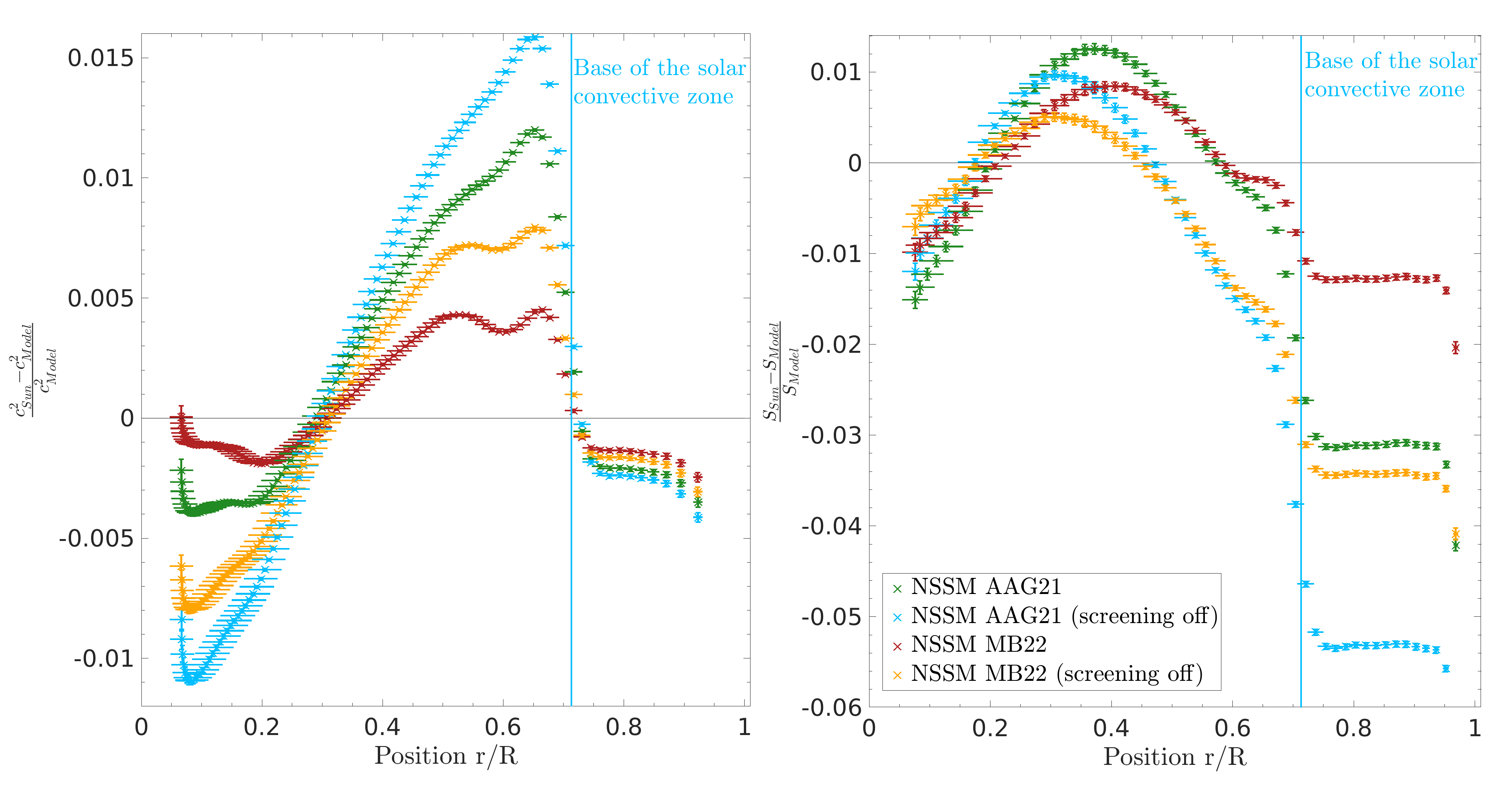}
	\caption{Left panel: Relative differences in squared adiabatic sound speed profile $(\rm{c}^{2})$ between the Sun and various NSSMs as a function of normalized radius, as determined from an SOLA inversion of helioseismic data. Right panel: Relative differences in entropy proxy profile $(\rm{S})$ between the Sun and various NSSMs as a function of normalized radius, as determined from an SOLA inversion of helioseismic data.}
		\label{Fig:c2NSSMScreen}
\end{figure*}

In addition to the effects on SSMs, we also compute helioseismic inversions of the squared adiabatic sound speed and the entropy proxy for models reproducing the lithium and beryllium depletion of the Sun, thus NSSMs using both MB22 and AAG21 abundances as well as testing the effect of a removal of the screening effects, following the suggestions of \citet{Mussack2011A, Mussack2011B}. The inversion results for these models for both sound speed and entropy proxy profiles are illustrated in Fig. \ref{Fig:c2NSSMScreen}. 

The results in this case are very clear, both from the variations of $\rho_{\rm{C}}$ and $\rm{T_{C}}$ and the sound speed and entropy proxy profiles. Including additional macroscopic mixing at the BCZ to reproduce the lithium and beryllium depletion slightly improves the sound speed profile at the BCZ at the expense of the core regions. As seen in previous works, the entropy proxy plateau is shifted away from the solar value when additional mixing is considered. In the case of neglecting the screening of nuclear reactions, this leads to a significant degradation of the agreement of all models with helioseismic constraints. Such increased discrepancies were already observed in \citet{Deal2025} and seem to speak against simply switching off screening as done in this work. 

Overall, from both the results presented in Figs. and \ref{Fig:C2SSM} and \ref{Fig:c2NSSMScreen}, we can conclude that the effects of the SFIII nuclear reaction rates are significant from a helioseismic point of view. Their inclusion is somewhat similar to a switch from the OPAL to the OPLIB opacities, with the agreement being improved in some regions while being degraded in others. This emphasizes the complexity of the solar problem, as the agreement with the precise and accurate constraints is significantly influenced by any change in physical ingredients of the models, it being the opacities \citep[see e.g.][]{Buldgen2017A,Buldgen2017A}, the transport of chemicals \citep[As discussed in the context of abundance revisions in][]{Buldgen2023,Buldgen2025Be} or the reaction rates, as shown here. 

\subsection{Neutrino fluxes}\label{Sec:NeutrinosModels}

\begin{figure}
	\centering
		\includegraphics[width=8.5cm]{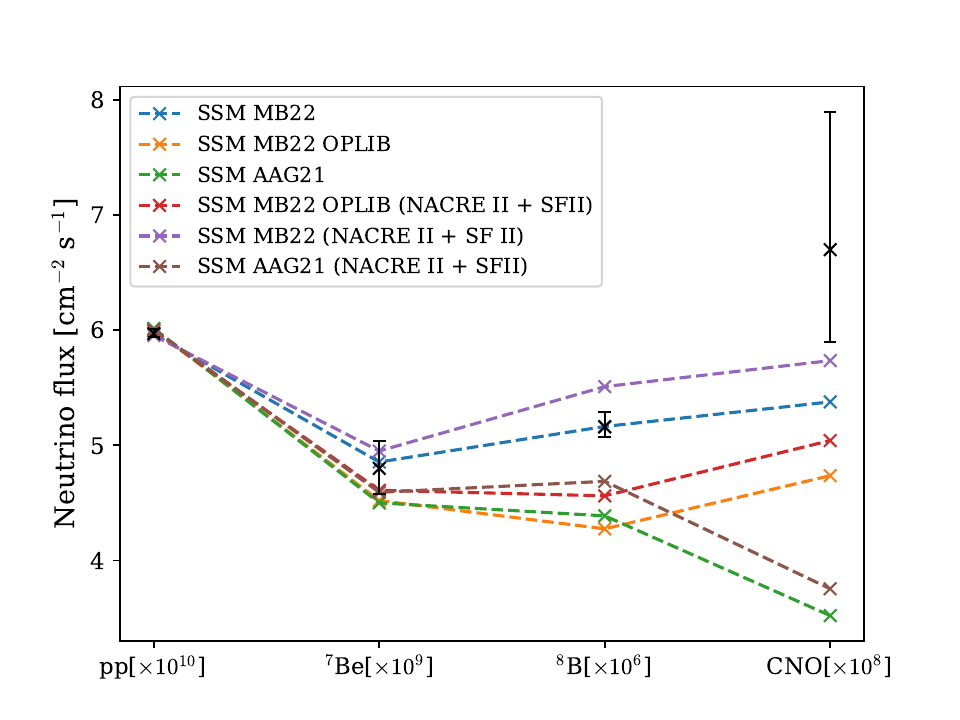}
	\caption{Predicted neutrino fluxes (pp, $^{7}\rm{Be}$, $^{8}\rm{B}$, CNO) by various SSMs in Table \ref{tabModels}, using various abundances, opacities and nuclear reaction rates. The effect of the SFIII reaction rates is clearly visible for various abundance tables.}
		\label{Fig:NeutrinosSSM}
\end{figure}

\begin{figure}
	\centering
		\includegraphics[width=8.5cm]{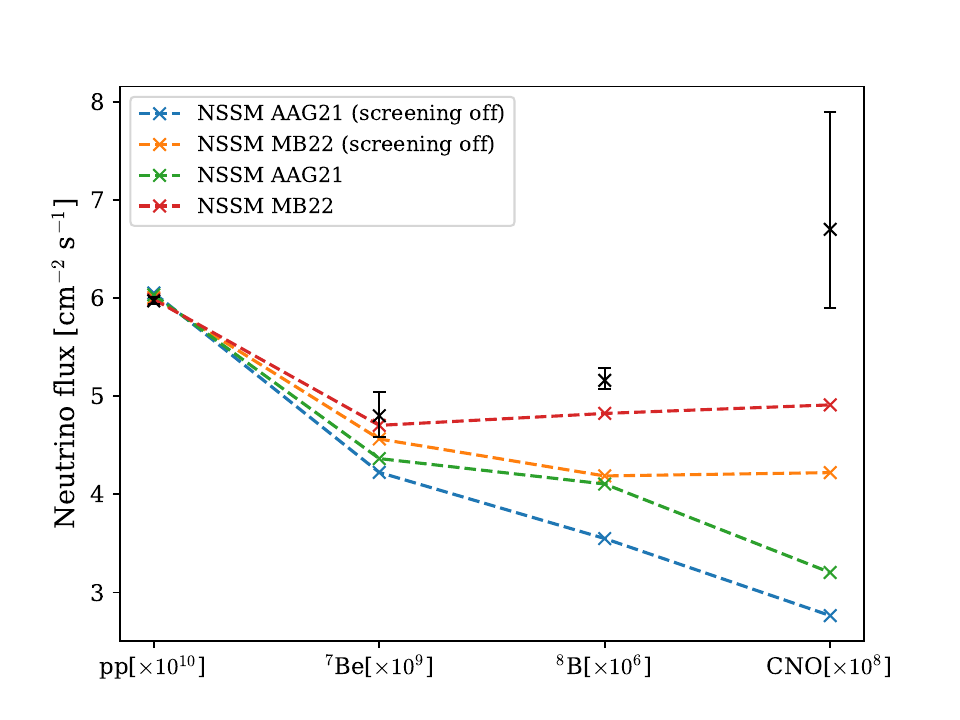}
	\caption{Predicted neutrino fluxes (pp, $^{7}\rm{Be}$, $^{8}\rm{B}$, CNO) by various NSSMs including macroscopic mixing in Table \ref{tabModels}, using various abundances, screening prescriptions and nuclear reaction rates.}
		\label{Fig:NeutrinosNSSM}
\end{figure}

While helioseismic constraints are clearly impacted by the changes of nuclear reaction rates, neutrino fluxes are also direct tracers of core conditions and may help us better analyze the impact of the SFIII rates on solar models. We illustrate in Figs. \ref{Fig:NeutrinosSSM} and \ref{Fig:NeutrinosNSSM} the predicted neutrino fluxes by  some of the models of Table \ref{tabModels} discussed in Sect. \ref{Sec:HelioModels}. The measured fluxes and the associated uncertainties were taken from \citet{OrebiGann2021}, except for the CNO neutrino flux that is taken from \citet{Basilico2023}.

From Fig. \ref{Fig:NeutrinosSSM}, we can clearly see that using the SFIII reaction rates leads to an overall reduction of the beryllium, $\phi_{\rm{Be}}$, boron, $\phi_{\rm{B}}$ and CNO neutrino fluxes, $\phi_{\rm{CNO}}$. The shifts are significant and lead to a worsening of the agreement for the AAG21 SSMs. For the MB22 SSMs, $\phi_{\rm{Be}}$ and $\phi_{\rm{B}}$ remain within the $1\sigma$ uncertainties, although one may see that the $\phi_{\rm{B}}$ is now in disagreement with the Borexino values at $\phi_{\rm{B}}=5.68^{+0.39}_{-0.41}\times 10^{6} \rm{cm}^{-2} \rm{s}^{-1}$ \citep{Borexino2018}. If we combine the SFIII reaction rates with the OPLIB opacities, then the situation is even worse, leading to strong disagreements with $\phi_{\rm{Be}}$, $\phi_{\rm{B}}$ and $\phi_{\rm{CNO}}$, even if the MB22 abundances are considered. This confirms the observations made by \citet{BuldgenS}, that the revision of the solar opacities lead to a worsening of the agreements of solar models with observational constraints. 

If we look at the results of NSSMs in Fig. \ref{Fig:NeutrinosNSSM}, the results are in line with expectations, with NSSMs including mixing leading to a lower core metallicity and thus larger disagreements with $\phi_{\rm{Be}}$, $\phi_{\rm{B}}$ and $\phi_{\rm{CNO}}$. We do not illustrate the results from a NSSM including macroscopic mixing computed with the SFIII rates and the OPLIB opacities, but one can easily see that the agreement would even worsen. Similarly, removing the electronic screening effect of the nuclear reaction rates has a significant impact on the neutrino fluxes. Contrary to what was observed in \citet{Deal2025}, here the calibrated models predict much lower beryllium, boron and CNO neutrino fluxes, in strong disagreements with measurements. This is due to the fact that the correction was only applied to reactions of the pp chain in \citet{Deal2025} and not to the CNO cycle, whereas here the screening coefficient of all reactions (pp and CNO) is simply set to 1.

The computations of neutrino fluxes for the SSM and NSSM models have shown a more complicated situation than what would have been seen from helioseismic constraints alone. While the SFIII reaction rates led to mild changes in the agreement of solar models with helioseismic inversions, the situation regarding neutrino fluxes is far more clear cut. Using the SFIII reaction rates in solar models leads to a reduction of $\phi_{\rm{Be}}$, $\phi_{\rm{B}}$ and $\phi_{\rm{CNO}}$. Combining these rates with OPLIB opacities or macroscopic mixing further worsen the agreement with neutrino measurements. In this respect, it is evident that neutrino flux measurement play a key role in constraining solar models and show that what was a somewhat unclear result from a helioseismic point of view is clearly seen as a degradation of the agreement of solar models from a neutrino flux point of view. These results are mostly due to the higher efficiency of the SFIII reaction rates for the pp chain compared to the NACRE II and SFII combinations used in our previous models. 

\section{Impact of the uncertainties on nuclear reaction rates}\label{Sec:UncertaintyNuclear}

As illustrated in Sect. \ref{Sec:Models}, changing nuclear reaction rates in solar calibrated models has a significant impact on both helioseismic constraints and neutrino fluxes. From a solar modelling perspective, this means that the nuclear reactions in solar models have some degree of uncertainty. Indeed, reference publications \citep{Villante2021} quote some uncertainties on key reactions and in this section, we will investigate the impact of some key reactions on helioseismic and neutrino constraints. We focus here on two reactions
\begin{align}
^{7}\rm{Be} + \rm{p} &\rightarrow \rm{^{8}B}+ \gamma, \label{eq:Boron}\\
^{14}\rm{N} + \rm{p} &\rightarrow \rm{^{15}O} + \gamma. \label{eq:CNO}
\end{align}
The first one is directly followed by disintegration of $^{8}\rm{B}$ into $^{8}\rm{Be}^{*}$ that is responsible for $\phi_{\rm{B}}$ while the second one is the so-called ``bottleneck reaction'' of the CNO cycle and thus directly impacts $\phi_{\rm{CNO}}$. Each reaction has its own uncertainties, with \citet{Acharya2025} reporting $3.41\%$ for reaction \ref{eq:Boron} and $8.3\%$ for reaction \ref{eq:CNO}. Higher values, around $10\%$ and $25\%$ can be found in \citet{Villante2021} following the recommendations of \citet{Adelberger2011} and \citet{Marta2011}.

\begin{figure*}
	\centering
		\includegraphics[width=15cm]{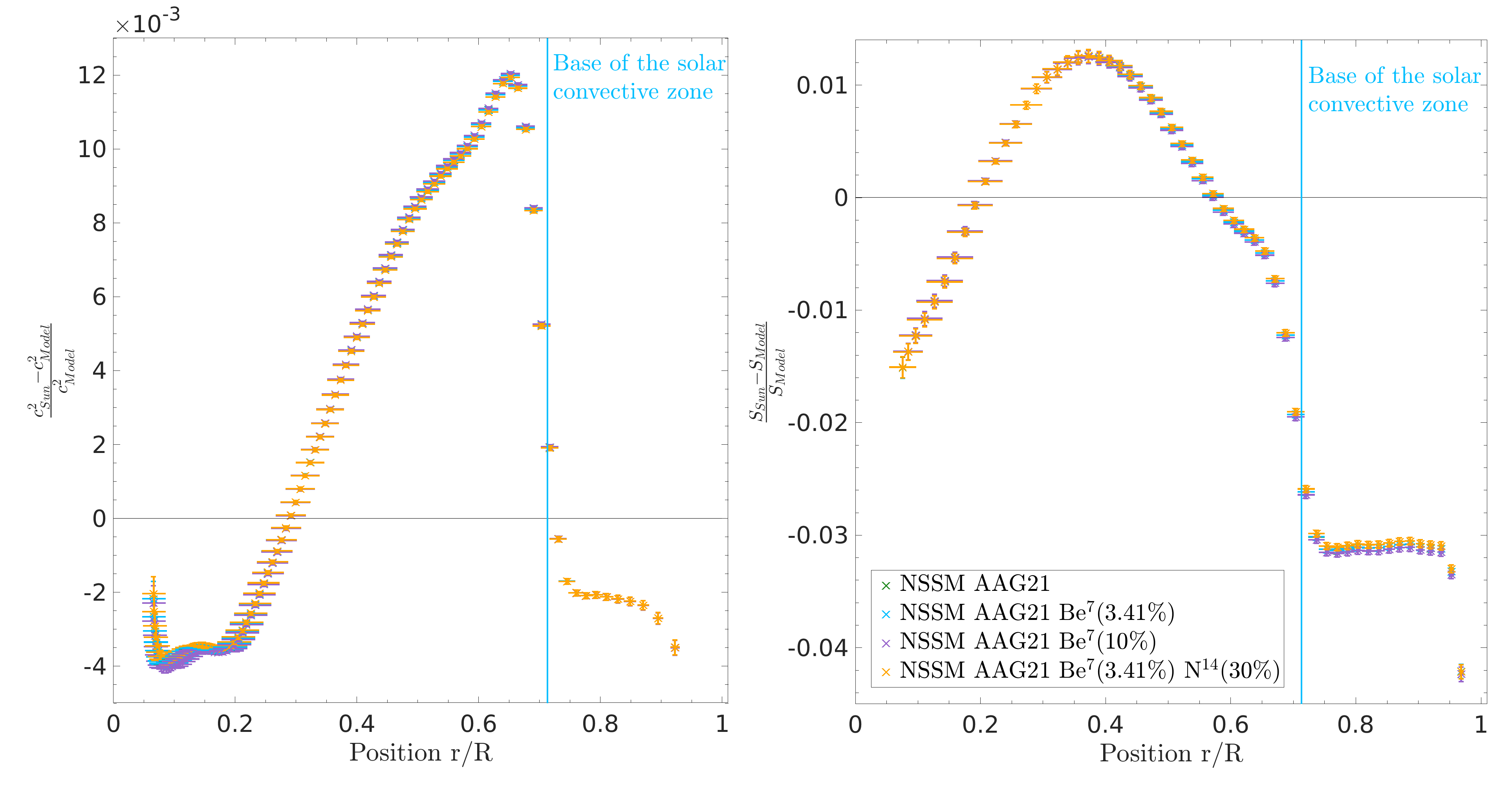}
	\caption{Left panel: Relative differences in squared adiabatic sound speed profile $(\rm{c}^{2})$ between the Sun and various NSSMs using the AAG21 abundances, including macroscopic mixing and modifications to both the $^{7}\rm{Be}(\rm{p}, \gamma)^{8}\rm{B}$ and $^{14}\rm{N}(\rm{p},\gamma)^{15}\rm{O}$ reaction rates.  as a function of normalized radius, as determined from an SOLA inversion of helioseismic data. Right panel: Same as the left panel but for relative differences in entropy proxy profile $(S)$ as a function of normalized radius.}
		\label{Fig:c2N14}
\end{figure*}

We start our analysis from a NSSM using the AAG21 abundances, the OPAL opacities and the SFIII reaction rates. We illustrate in Fig. \ref{Fig:c2N14} the impact of considering an increase of $3.41\%$ and $10\%$ respectively of the $^{7}\rm{Be}(\rm{p}, \gamma)^{8}\rm{B}$ reaction as well as the combined impact of a $3.41\%$ increase in the $^{7}\rm{Be}(\rm{p}, \gamma)^{8}B$ reaction and a $30\%$ increase of the $^{14}\rm{N}(\rm{p},\gamma)^{15}\rm{O}$ reaction rate, on helioseismic constraints. As can be seen, the impact of such modifications to the nuclear reaction rates are completely negligible on helioseismic constraints. This is also seen in Table \ref{tabModels} and is a direct consequence of the minor role of both reactions in the total energy budget of the Sun \citep[see][and refs therein for a discussion]{Villante2021}. Therefore, helioseismic constraints are unable to provide detailed information on specific reaction rates and we must rely on solar neutrinos as the only source of information. 

\subsection{The $^{14}\rm{N}(\rm{p},\gamma)^{15}\rm{O}$ reaction}

\begin{figure}
	\centering
		\includegraphics[width=8.5cm]{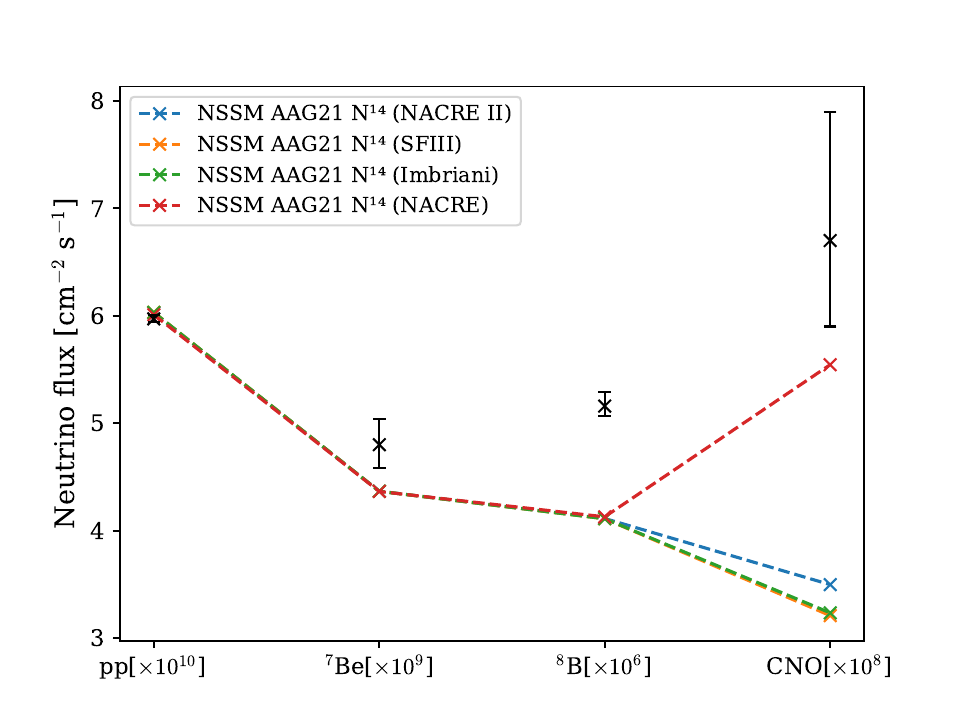}
	\caption{Predicted neutrino fluxes (pp, $^{7}\rm{Be}$, $^{8}\rm{B}$, CNO) by various NSSMs including macroscopic mixing in Table \ref{tabModels}, using the AAG21 abundances and various references for the $^{14}\rm{N}(\rm{p},\gamma)^{15}\rm{O}$.}
		\label{Fig:NeutrinosN14}
\end{figure}

To put things in perspective regarding the $^{14}\rm{N}(\rm{p},\gamma)^{15}\rm{O}$, we first consider the impact of varying the $^{14}\rm{N}(\rm{p},\gamma)^{15}\rm{O}$ reaction amongst the known references in the literature. We illustrate in Fig. \ref{Fig:NeutrinosN14} the predicted neutrino fluxes by the various solar models using the NACRE \citep{Angulo1999}, the \citet{Imbriani2005}, the NACRE II \citep{Xu2013} and SFIII rate for the $^{14}\rm{N}(\rm{p},\gamma)^{15}\rm{O}$ reaction, while all other reactions in these solar calibrations have been fixed to the SFIII reaction rates.

From these results, we can see that, as expected, the only impacted neutrino flux is $\phi_{\rm{CNO}}$. We can also comment on the strong impact of the revision of the NACRE reaction rate and the consistency between the results from NACRE II, SFIII and those from \citet{Imbriani2005}. However, looking at Figs. \ref{Fig:NeutrinosSSM} and \ref{Fig:NeutrinosNSSM}, it is also clear that the $^{14}\rm{N}(\rm{p},\gamma)^{15}\rm{O}$ is only one ingredient of the issue and that other factors influencing the solar calibration procedure (such as the opacities, the transport of chemicals, ...) may have a direct impact on the fluxes through their impact on the overall solar core conditions. As such, the $^{14}\rm{N}(\rm{p},\gamma)^{15}\rm{O}$ is one key piece of the puzzle that will not impact the disagreements with other neutrino fluxes \citep{Frentz2022}. Another conclusion we may draw from Fig. \ref{Fig:NeutrinosN14} is that even extreme modifications such as those between NACRE and NACRE II results are not sufficient on their own to bring back the AAG21 NSSM models in agreement with the observed $\phi_{\rm{CNO}}$ at Borexino \citep{Borexino2020,Basilico2023}.

\subsection{Combined impact of both $^{7}\rm{Be}$, $^{8}\rm{B}$ and $^{14}\rm{N}(\rm{p},\gamma)^{15}\rm{O}$}

\begin{figure}
	\centering
		\includegraphics[width=8.5cm]{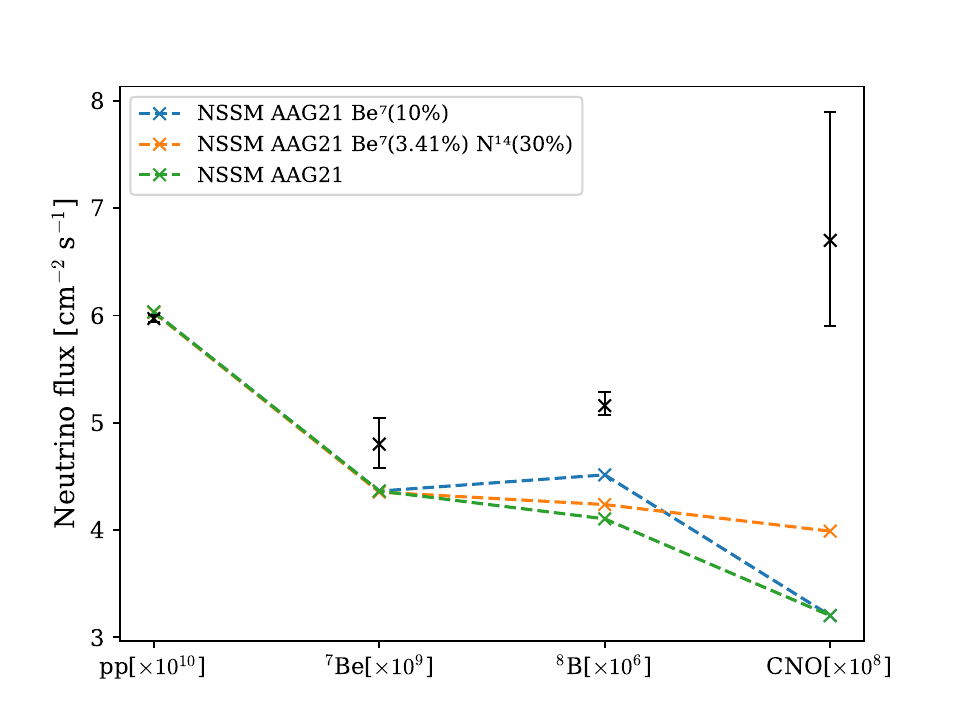}
	\caption{Predicted neutrino fluxes (pp, $^{7}\rm{Be}$, $^{8}\rm{B}$, CNO) by various NSSMs including macroscopic mixing in Table \ref{tabModels}, using the AAG21 abundances and including modifications to both the $^{7}\rm{Be}(\rm{p}, \gamma)^{8}B$ and $^{14}\rm{N}(\rm{p},\gamma)^{15}\rm{O}$ reaction rates.}
		\label{Fig:NeutrinosBe7N14}
\end{figure}

As shown in Fig. \ref{Fig:c2N14}, the $^{7}\rm{Be}(\rm{p}, \gamma)^{8}\rm{B}$ and $^{14}\rm{N}(\rm{p},\gamma)^{15}\rm{O}$ reaction rates have little impact on helioseismic constraints. The situation is completely different when looking at neutrino fluxes, as shown in Fig. \ref{Fig:NeutrinosBe7N14}. Variations of the $^{7}\rm{Be}(p, \gamma)^{8}B$ have a direct impact on $\phi_{\rm{B}}$, as well as a much more modest impact on $\phi_{\rm{Be}}$. We can however see that even a $10\%$ modification of the $^{7}\rm{Be}(\rm{p}, \gamma)^{8}\rm{B}$ is insufficient to bring back AAG21 models including macroscopic mixing in agreement with the $\phi_{\rm{B}}$ value from \citet{OrebiGann2021}. 

Similarly the combined $3.41\%$ uncertainties on the $^{7}\rm{Be}(\rm{p}, \gamma)^{8}\rm{B}$ reaction rate and $30\%$ on the 
$^{14}\rm{N}(\rm{p},\gamma)^{15}\rm{O}$ rate are largely insufficient to restore the agreement with both the measured $\phi_{\rm{B}}$ and $\phi_{\rm{CNO}}$. This is unsurprising, as the problem was already present for SSM with older nuclear rates \citep[see the discussion][]{Basilico2023}. 

\section{Discussion}\label{Sec:Discussion}

As illustrated in the previous sections, the use of the recently published SFIII nuclear reaction rates \citep{Acharya2025} instead of the NACRE II reaction rates does not translate in a systematic improvement of solar models with helioseismic and neutrino constraints. While the agreement with helioseismic constraints remain comparable or even slightly improved in some regions such as the solar core (at least as far as adiabatic sound speed and entropy proxy are considered), the agreement with neutrino fluxes worsens in a significant way. Including the SFIII reaction rates induces an overall reduction of the $\phi_{\rm{Be}}$, $\phi_{\rm{B}}$ and $\phi_{\rm{CNO}}$ fluxes. 

While one could agree that the $\phi_{\rm{Be}}$ and $\phi_{\rm{B}}$ reduction is not a drastic issue for the MB22 abundances (provided that one favours the $\phi_{\rm{B}}$ values from \citet{OrebiGann2021} over those of Borexino), this is only true when neglecting macroscopic mixing at the BCZ to reproduce light element depletion and favouring the OPAL opacity tables over the OPLIB tables. However, the situation is naturally worse for the AAG21 models which show even larger disagreements. We also find that modifications to electronic screening are not able to improve the situation in our study, as both helioseismic constraints and neutrino fluxes show strong discrepancies for calibrated models with such changes. These differences seem to directly come from lower efficiency of the reactions, as deactivating screening implies a higher potential barrier to be crossed. Compared to previous tests in \citet{Deal2025}, only the screening of the ppI reactions were set to one, meaning that in this case, the higher central temperature and metallicity directly impacted positively $\phi_{\rm{CNO}}$. 

The origin of the differences between SFIII and our older reaction rates is a bit more subtle. It can be studied when looking at the arguments of the integrals of the various neutrino fluxes which are directly proportional to the nuclear rates.
\begin{align}
\varphi_{\rm{pp}}(r) &= 4\pi r^2 \frac{1}{2}\rho^2 \mathcal{N}_1^2 N_A^2 \langle\sigma v\rangle_{11} \label{eq:phipp}, \\
\varphi_{^{7}\rm{Be}}(r) &= 4\pi r^2\rho^2\mathcal{N}_4\mathcal{N}_1 N_A^2\sqrt{\frac{\langle\sigma v\rangle_{11}}{2\langle\sigma v\rangle_{33}}}\langle\sigma v\rangle_{34},\label{eq:phiBe}\\
\varphi_{^8\rm{B}}(r) &= 4\pi r^2\rho^2 \mathcal{N}_7 \mathcal{N}_1 N_A^2 \langle\sigma v\rangle_{17},\label{eq:phiB}\\
\varphi_{\rm{CNO}} (r) &= 4\pi r^2 \rho^2 \mathcal{N}_1 \mathcal{N}_1 N_A^2(\mathcal{N}_{12} \langle\sigma v\rangle_{112} +  \mathcal{N}_{14} \langle\sigma v\rangle_{114} \nonumber \\ &+ \mathcal{N}_{16} \langle\sigma v\rangle_{116})\label{eq:phiCNO},
\end{align}
with $\mathcal{N_{X}}$ the abundance in $ \left[\rm{mol/g}\right]$ of the following elements: $1$ indicates $^{1}$H, $4$ is $^{4}$He, $7$ is $^{7}$Be, $14$ is $^{14}$N, $12$ is $^{12}$C and $16$ is $^{16}$O. $N_{A}$ is Avogadro's number, $\langle\sigma v\rangle_{XX}$ are the various average cross sections for the different reactions considered here, $\rho$ is the local density and $r$ the radial coordinate.

\begin{figure*}
	\centering
		\includegraphics[width=13cm, angle =270 ]{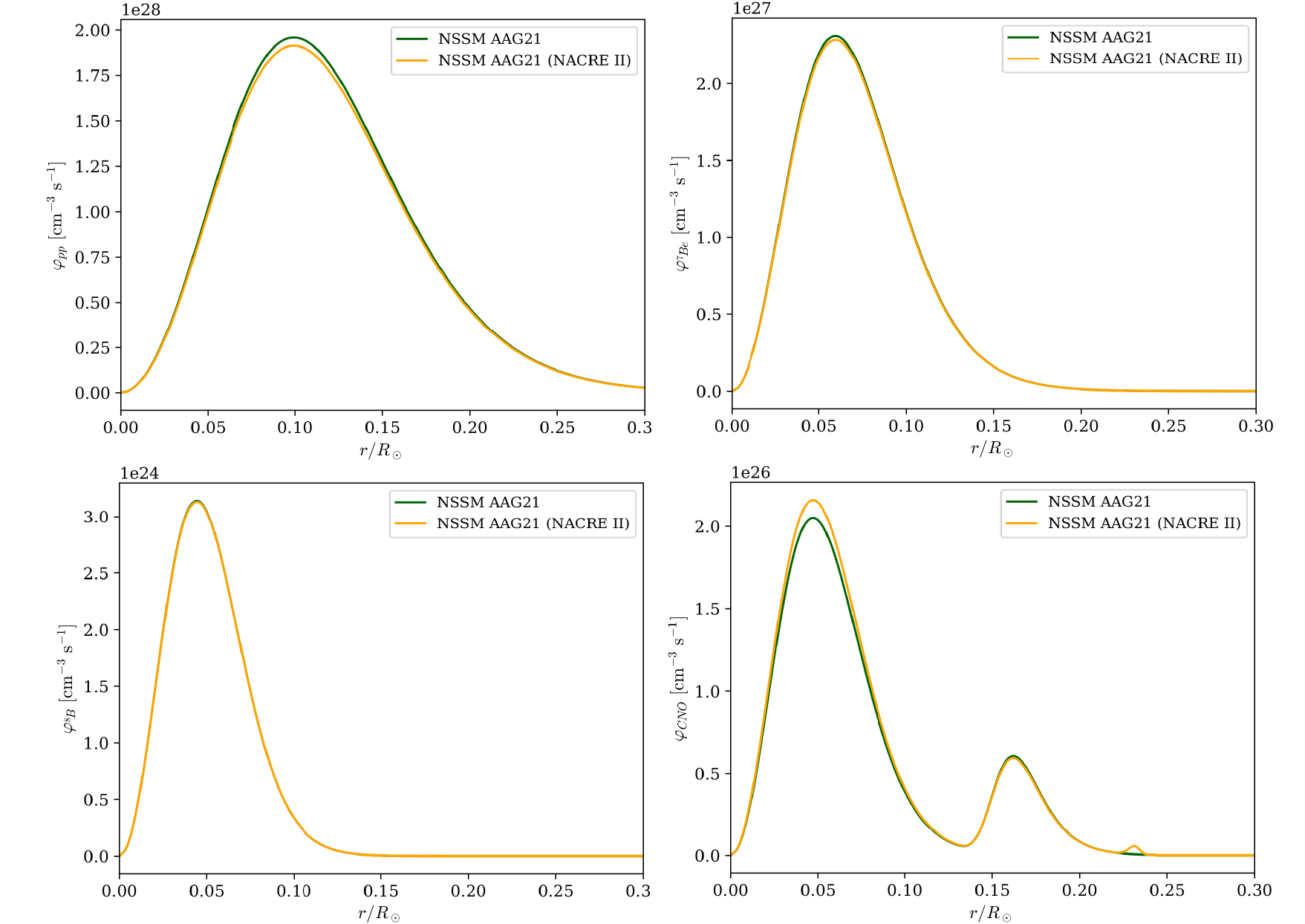}
	\caption{Arguments of the integrals described in Eqs. \ref{eq:phipp},\ref{eq:phiBe},\ref{eq:phiB} and \ref{eq:phiCNO} for the NSSM model using the AAG21 abundances and either the SFIII or the combination of NACRE II and SFII reaction rates used in our previous publications. Upper left panel: argument for the pp neutrinos. Upper right panel: argument for the $^{7}$Be neutrinos. Lower left panel: argument for the $^{8}$B neutrinos. Lower right panel: argument for the CNO neutrinos.}
		\label{Fig:Integ}
\end{figure*}

To better visualize what is happening, we plot in Fig. \ref{Fig:Integ} these argument of the integrals for $\phi_{\rm{pp}}$, $\phi_{\rm{Be}}$, $\phi_{\rm{B}}$ and $\phi_{\rm{CNO}}$ for fixed temperature and chemical composition (in other words, without recalibrating the solar models as this could induce compensation effects). We can see that using the same thermodynamic coordinates, the arguments of $\phi_{\rm{pp}}$ and $\phi_{\rm{Be}}$ are slightly higher for a fixed composition, while $\phi_{\rm{CNO}}$ is significantly lower, while that of $\phi_{\rm{B}}$ is essentially the same. These small changes allow to explain the slightly lower $\rm{X_{C}}$, $\rm{Z_{C}}$, $T_{C}$ and $\rho_{C}$ values. 

The slightly higher efficiency of the pp chain with SFIII requires to compensate by lowering $\rm{X_{C}}$ and $T_{C}$ which allow to reproduce both the solar luminosity at the solar age and $\phi_{\rm{pp}}$. The slightly lower $T_{C}$ leads to a lowering of $\phi_{\rm{Be}}$, partially compensated by the slightly higher reaction rate, while for $\phi_{\rm{B}}$ where the integrand is similar for the same thermodynamical conditions, the lowering of the temperature leads to a larger decrease of $\phi_{\rm{B}}$. The situation is even worse for $\phi_{\rm{CNO}}$ as a lower efficiency of the reaction is observed in Fig. \ref{Fig:Integ} at fixed conditions, while here both temperature and central metallicity effects lead to a lowering of the flux for the solar calibrated models with SFIII. These observations emphasize the extreme sensitivity of the neutrino fluxes to the reaction rates and the fact that a modification of a rate might actually trigger a trickle down effect, by changing the calibrated central temperature and chemical composition in a somewhat similar effect to what is observed when changing opacity tables in solar models.

\section{Conclusion}\label{Sec:Conc}

In this study, we have presented new standard and non-standard solar models, reproducing the solar light element depletion, computed with the latest SFIII nuclear reaction rates \citep{Acharya2025}. We started in Sect. \ref{Sec:Models} by presenting the global properties of the models, before looking into the results of helioseismic inversions in Sect. \ref{Sec:HelioModels} and neutrino fluxes in Sect. \ref{Sec:NeutrinosModels}. While we find that using the SFIII reaction rates overall improves slightly the agreement of solar models with observations, the same is not true for neutrino fluxes that see a significant decrease in $\phi_{\rm{Be}}$ and $\phi_{\rm{B}}$ and $\phi_{\rm{CNO}}$. The origin of the enhanced discrepancies stems from both the alteration of the thermodynamical conditions of the core of the calibrated models and for the CNO fluxes, of a slightly reduced efficiency of the cycle. These results remain valid for various assumed solar abundances and opacity tables.

In addition to presenting the new models computed with the SFIII rates, we also investigated the impact of fully neglecting screening of all nuclear reactions. In this case, fully deactivating the electronic screening for all reactions (both in the pp chains and the CNO cycle) leads to a significant decrease of neutrino fluxes. This emphasizes the importance of determining the potential need for dynamical corrections to electronic screening that have so far been neglected in the literature. 

The last tests we performed are linked with the existing uncertainties on key nuclear reactions, presented in Sect. \ref{Sec:UncertaintyNuclear}. From this analysis, we find that the current uncertainties on two key reactions linked with neutrino fluxes, the $^{7}\rm{Be}(\rm{p}, \gamma)^{8}\rm{B}$ reaction and the $^{14}\rm{N}(\rm{p},\gamma)^{15}\rm{O}$ reaction, are insufficient to reconcile models with observed neutrino fluxes. This is particularly acute for models using the OPLIB opacities or NSSMs matching the lithium and beryllium depletion of the Sun. 

Overall, the recent SFIII nuclear reaction rates appear to have a similar impact on the solar problem than the publication of the new opacity tables \citep{LePennec,Mondet,Colgan}. While they improve the agreement with some constraints, they also degrade it with other observations, leaving further revision of physical ingredients, including nuclear reaction rates \citep{Villante2021} as paramount. In addition, the impact of planetary formation \citep{Kunitomo2021,Kunitomo2022,Kunitomo2025} or mass-loss \citep[See e.g.][]{Guzik2010,Zhang2019} should not be neglected as a potential source of further contrast between the surface and core metallicity of the Sun that would bring the solar models in agreement with the observed CNO neutrino flux of Borexino. 

\begin{acknowledgements}
We thank the referee for their careful reading of the
manuscript. GB acknowledges fundings from the Fonds National de la Recherche Scientifique (FNRS) as a postdoctoral researcher. 

\end{acknowledgements}

\bibliography{biblioarticleSFIII}

\end{document}